\pdfoutput=1
\pdfoutput=1
\pdfoutput=1
\pdfoutput=1
\pdfoutput=1
\RequirePackage{fix-cm} 
\documentclass[twocolumn,epjc3-upd,fleqn]{svjour3}
\smartqed  % flush right qed marks, e.g. at end of proof
\RequirePackage{graphicx}
\RequirePackage{subfigure}
\usepackage{rotating}
\usepackage{amsmath}
\usepackage{mathtools}
\usepackage{multirow}
\usepackage{cite}
\usepackage[switch]{lineno} 
\RequirePackage[colorlinks,citecolor=blue,urlcolor=blue,linkcolor=blue]{hyperref}
\usepackage{doi}
\usepackage[]{units}
\usepackage{xspace}
\journalname{Eur. Phys. J. C}
\def\teb{2.7} %limit
\def\geb{2.56} %limit
\def\bestgeb{(-0.06 \pm 1.37)}
\def\tee{2.6} %limit
\def\gee{2.69} %limit

\def\vita-dim{$T_{1/2}$\xspace}

\def\be{\begin{equation}}
\def\ee{\end{equation}}

\begin{document}
%\linenumbers
\title{Search for double $\beta$-decay modes of $^{64}$Zn using purified zinc}

\author{F.~Bellini\thanksref{Roma,INFNRoma}
\and M.~Beretta\thanksref{MIB,INFNMiB,UCB}
\and L.~Cardani\thanksref{INFNRoma}
\and P.~Carniti\thanksref{MIB,INFNMiB}
\and N.~Casali\thanksref{INFNRoma}
\and E.~Celi\thanksref{GSSI,LNGS,e1}
\and D.~Chiesa\thanksref{MIB,INFNMiB}
\and M.~Clemenza\thanksref{MIB,INFNMiB}
\and I.~Dafinei\thanksref{INFNRoma}
\and S.~Di~Domizio\thanksref{Genova,INFNGenova}
\and F.~Ferroni\thanksref{GSSI,INFNRoma}
\and L.~Gironi\thanksref{MIB,INFNMiB}
\and Yu.V.~Gorbenko\thanksref{KIPT}
\and C.~Gotti\thanksref{MIB,INFNMiB}
\and G.P.~Kovtun\thanksref{KIPT,KAR}
\and M.~Laubenstein\thanksref{LNGS}
\and S.~Nagorny\thanksref{LNGS,GSSI,QUEEN,e2}
\and S.~Nisi\thanksref{LNGS}
\and L.~Pagnanini\thanksref{GSSI,LNGS,e3}
\and L.~Pattavina\thanksref{LNGS}
\and G.~Pessina\thanksref{INFNMiB}
\and S.~Pirro\thanksref{LNGS}
\and E.~Previtali\thanksref{MIB,INFNMiB}
\and C.~Rusconi\thanksref{USC,LNGS}
\and K.~Sch\"affner\thanksref{GSSI,LNGS}
\and A.P.~Shcherban\thanksref{KIPT}
\and D.A.~Solopikhin\thanksref{KIPT}
\and V.D.~Virich\thanksref{KIPT}
\and C.~Tomei\thanksref{INFNRoma}
\and M.~Vignati\thanksref{INFNRoma}
}

\institute{Dipartimento di Fisica, Sapienza Universit\`{a} di Roma, Roma I-00185 - Italy \label{Roma}
\and
INFN - Sezione di Roma, Roma I-00185 - Italy\label{INFNRoma}
\and
Dipartimento di Fisica, Universit\`{a} di Milano - Bicocca, Milano I-20126 - Italy\label{MIB}
\and
INFN - Sezione di Milano - Bicocca, Milano I-20126 - Italy\label{INFNMiB}
\and
Gran Sasso Science Institute, I-67100, L'Aquila - Italy\label{GSSI}
\and
INFN - Laboratori Nazionali del Gran Sasso, Assergi (L'Aquila) I-67100 - Italy\label{LNGS}
\and
Dipartimento di Fisica, Universit\`{a} di Genova, Genova I-16146 - Italy\label{Genova}
\and
INFN - Sezione di Genova, Genova I-16146 - Italy\label{INFNGenova}
\and
National Science Center “Kharkov Institute of Physics and Technology”, 61108 Kharkov, Ukraine\label{KIPT}
\and
Karazin Kharkov National University, Kharkov 61022, Ukraine \label{KAR}
\and
Department of Physics and Astronomy, University of South Carolina, Columbia, SC 29208 - USA\label{USC}
\and
Present Address: University of California, Berkeley, CA 94720, US\label{UCB}
\and
Present Address: Queen's University, Physics Department, K7L 3N6, Kingston (ON), Canada \label{QUEEN}
}

\thankstext{e1}{email: \href{mailto:emanuela.celi@gssi.it}{emanuela.celi@gssi.it}}
\thankstext{e2}{email: \href{mailto:sn65@queensu.ca}{sn65@queensu.ca}}
\thankstext{e3}{email: \href{mailto:lorenzo.pagnanini@gssi.it}{lorenzo.pagnanini@gssi.it}}

\date{Received:  / Accepted: }
\twocolumn
\maketitle
\begin{abstract}
The production of ultra-pure raw material is a crucial step to ensure the required background level in rare event searches. In this work, we establish an innovative technique developed to produce high-purity (99.999\%) granular zinc. We demonstrate the effectiveness of the refining procedure by measuring the internal contaminations of the purified zinc with a high-purity germanium detector at the Laboratori Nazionali del Gran Sasso. The total activity of cosmogenic activated nuclides is measured at the level of a few mBq/kg, as well as limits on naturally occurring radionuclides are set to less than mBq/kg. The excellent radiopurity of the zinc sample allows us to search for electron capture with positron emission
and neutrinoless double electron capture of $^{64}$Zn, setting the currently most stringent lower limits on their half-lives, $T_{1/2}^{\varepsilon\beta^+} > \teb \times 10^{21}~\text{yr~}$ (90\% C.I.), and $T_{1/2}^{0\nu2\varepsilon}> \tee \times 10^{21}~\text{yr~}$ (90\% C.I.), respectively.
\end{abstract}

\section{Introduction}
\label{Sec:Introduction}
The search for rare events such as dark matter interactions \cite{PhysRevD.98.030001} or neutrinoless double-beta decay \cite{Dolinski:2019nrj} has an important role in the investigation of physical processes not predicted by the standard model. The common experimental effort for next-generation experiments focuses on the reduction of detector contamination, to reach an extremely low background index. In this scenario, the purification of the initial materials used for the detector construction acquires a pivotal relevance. 
Zinc-containing crystals are widely used in rare event physics as low-temperature calorimeters (ZnSe in CUPID-0 \cite{Azzolini:2018tum}, ZnMoO$_4$ in LUMINEU \cite{Armengaud:2017hit}), room-temperature scintillators (ZnWO$_4$ \cite{Barabash_2020,BARABASH201677,BELLI201989}) or semiconductor detectors (CdZnTe in COBRA \cite{Ebert:2015rda}). 
To understand the importance of the purity of the initial materials, we consider the specific case of ZnSe \cite{Dafinei:2017xpc}. The most common production method for these crystals is the growth by melt crystallization under high inert gas pressure, known as Bridgman-Stockbarger technique \cite{BStech}. With this method large volume crystals can be grown, reaching up to 60 mm in diameter and 2 kg in mass. The downside of this technique is the lack of perfection in the produced crystalline structure. During crystallization, a fraction of the ZnSe compound in the liquid phase dissociates, transferring part of the single components (Zn and Se) to the cold zone. The transfer is due either to diffusion through the semi-penetrable graphite crucible walls, or to evaporation from the free surface of the melt. Because of differences in diffusion coefficients, the melt results enriched in one of the components, causing deviations from the stoichiometry that can be significantly larger than 1\%. Such a process leads to the formation of intrinsic point defects (IPD) in the crystalline structure, affecting both the scintillating and bolometric crystal properties, as well as the detector radiopurity.
Since this deviation from stoichiometry increases together with the contamination of the raw initial materials, their chemical purity impacts directly on the detector performance. This is true in particular for bolometric detectors, where defects of the crystalline structure affect the phonon scattering in the crystal. As a result the thermal conductivity of the detector decreases, causing a loss in energy resolution. It is therefore mandatory to use initial materials with superior chemical and radio-purity to obtain high-performance scintillating bolometers based on zinc-containing crystals.
In this work, we introduce a novel method for zinc purification
(Sec. \ref{Sec:purification}), validating the whole production through mass spectrometry (Sec.~\ref{Sec:refinement}) and $\gamma$-spectroscopy (Sec.~\ref{Sec:hpge}). The excellent radiopurity achieved allows us to search for double $\beta$-decay of $^{64}$Zn, in particular neutrinoless and two-neutrino modes of electron capture with positron emission ($\varepsilon\beta^{+}$), as well as neutrinoless double electron capture ($0\nu2\varepsilon$), setting the most stringent lower limits on their half-lives to date (Sec.~\ref{Sec:dbd}).

\section{A new process for zinc purification}
\label{Sec:purification}
\begin{figure}
    \centering
    \includegraphics[width=0.5\textwidth]{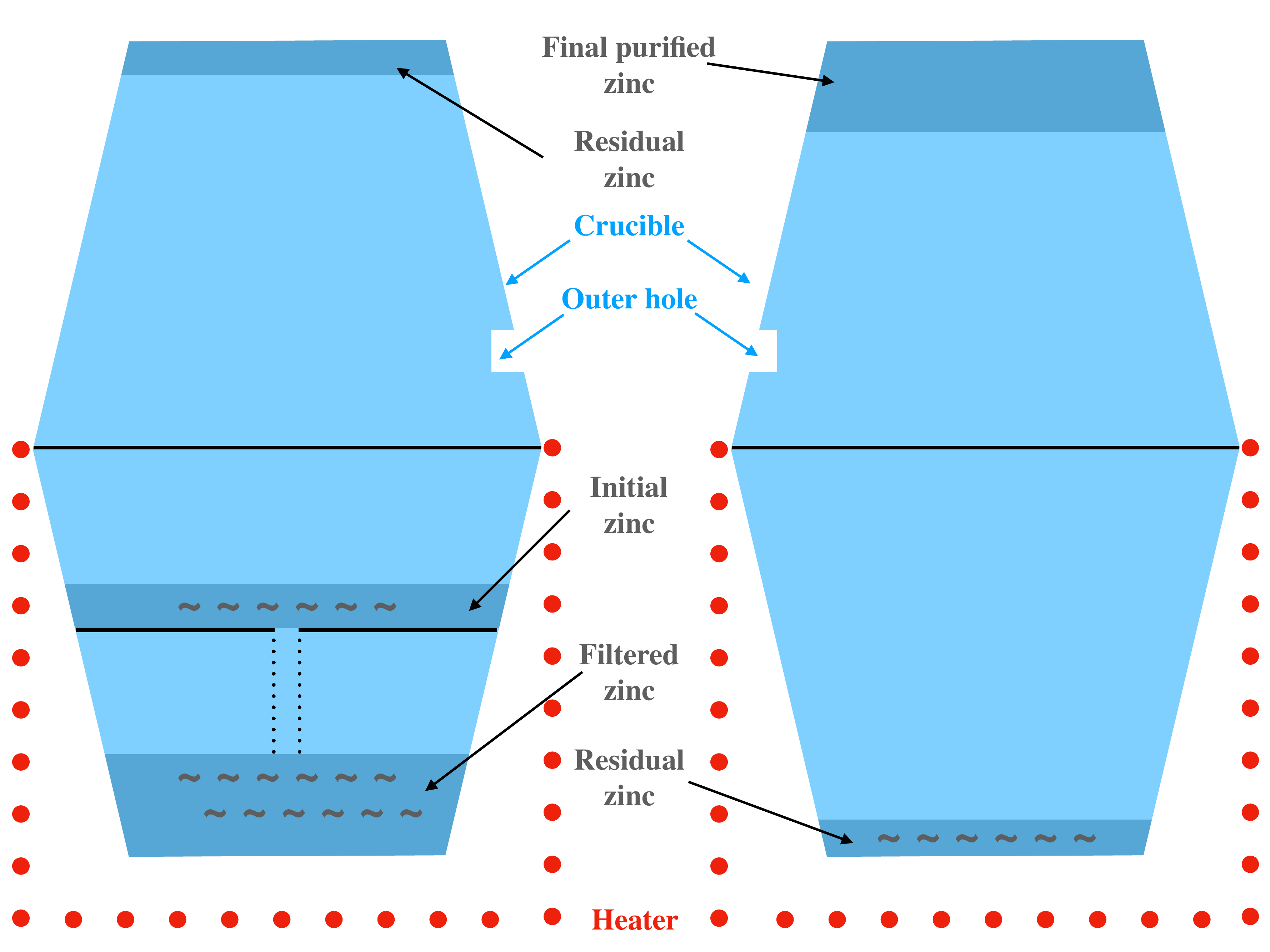}
    \caption{Scheme of the set-up for zinc purification: a) stage of filtration and refining against of the volatile impurities (left); b) stage of purification against of the non-volatile impurities (right). We melt the initial zinc using a heater, obtaining the filtered zinc. We remove most of the volatile impurities through the  outlet hole, while a small amount of them remains in the thin layer of condensed zinc, also removed. By further heating the sample, all zinc evaporates, forming a layer of purified zinc, while non-volatile impurities form a residue in the crucible.}
    \label{fig:purification}
\end{figure}

The raw zinc, chosen with purity level higher than 99.98\%, is prepared for the purification process trough the following steps: (i) cutting of initial metal ingot in order to get pieces of about 2 kg of mass, (ii) etching of Zn pieces in the nitric acid solution (0.1 M HNO$_3$), (iii) washing of etched Zn pieces in HP water, (iv) drying of washed Zn pieces, and (v) sampling for tracing of contamination level by means of mass-spectrometry. The actual purification process presented here consists then of two phases: (i) the refining of the initial zinc by filtration and distillation under vacuum, and (ii) the formation of granules of (3 - 5) mm in diameter by dripping of molten metal in the coolant agent. This last step modifies the material in a suitable form for the subsequent stages of crystal production.
The innovative method used to implement the first step has been developed at the National Science Center ``Kharkov Institute of Physics and Technology" (Kharkov, Ukraine). The procedure involves the filtration  of the molten metal and its subsequent distillation by condensing the metal steam into the solid phase \cite{Kovtun:2011}. The distillation process is carried out at a temperature $T_D=T_M+(40 - 50)$~$^{o}$C, while the condensation process at a temperature $T_C=T_M-(30 - 40)$~$^{o}$C, being T$_M= 419~^{o}$C the melting temperature. 
The distillation set-up is made from high-purity high-density graphite, characterized by chemical inertness to zinc and a minimal content of impurities. To further increase the graphite purity, the setup was thermally treated under vacuum at (1000 - 1100) $^{o}$C before use. The total concentration of all possible contaminants in graphite has been therefore reduced below 1.0 ppm. The distillation set-up was assembled of two identical pieces, as shown in Fig~\ref{fig:purification}: the bottom one serves as a crucible, while the upper part is a collector for the condensed Zn metal. The small outlet hole in the lateral wall of the condenser is intended for evacuation in order to remove volatile impurities. The crucible is heated by the heater, while the condenser is heated indirectly by the thermal radiation of the crucible, as well as by the heat transfer from the stream of the distillable metal. The crucible volume was designed to hold 2.0 kg of loaded initial zinc. The operating pressure during of the zinc refining was about $(10^{-3}-10^{-2})$~Pa.
At the first stage, the purification process was carried out against volatile impurities (like a Na, K etc.) by their condensation on the inner surface of the condenser (see Fig~\ref{fig:purification}, left). A part of them is also removed continuously pumping the crucible volume through the outer hole. In order to remove oxides and other slags, the molten metal was filtered, through a plate with a small conicity and a hole in the middle. At the end of this stage, a certain amount of zinc metal that contains a high concentration of the volatile impurities is removed from the coldest part of the condenser.
The second stage is carried out in order to remove the non-volatile impurities (like a Cu, Fe, Si, Ni, Co, V, Cr, Al, Tl, Bi, Mn etc.) by distillation of the zinc metal, poured in the crucible after filtration. After evaporation (up to 95\% of loaded material), the pure zinc is condensed on the inner part of the condenser, while the non-volatile impurities remain as residue on the bottom of the crucible (see Fig~\ref{fig:purification}, right). Such combination of refining stages significantly improves the efficiency of the whole purification process and product yield, which is better than 95\% from the initial loaded zinc.
At the end of this process, the purified zinc was sent for the granulation in a dedicated device by dripping of the molten metal into high-purity water used as a cooling agent\footnote{Patent on metal granulation N. 131214 Ukraine - Publ. 01/10/2019 - Bull. N. 1 - By A.P Shcherban, Yu.V. Gorbenko, G.P. Kovtun, D.A. Solopikhin.}. 
The final high-purity zinc was obtained in form of granules with 3--4 mm in diameter resulting in a total mass of 15 kg. The yield of the granulation process is about 99\%.

\section{Results of refinement}
\label{Sec:refinement}
We did a general comparative analysis of elemental impurities in the zinc metal before and after purification by combining the Laser (LMS) and Inductively Coupled Plasma (ICP) Mass-Spectrometry methods. While chemical purity of initial zinc metal was studied at NSC KIPT (Kharkov, Ukraine) using a High-Resolution Double-Focusing Laser Mass-Spectrometer EMAL-2 by Mattauch-Herzog, the residual contamination of purified metal was examined with the help of High-Resolution Inductively Coupled Plasma-Mass Spectrometric analysis (HR-ICP-MS, Thermo Fisher Scientific ELEMENT2) at the Gran Sasso Laboratory (Assergi, Italy). To reduce cosmogenic activation, we have shipped the zinc from Ukraine to Italy by land. The results of comparative measurements are listed in Tab.~\ref{tab:1}.
\begin{table}
\centering
\caption{Impurity concentration in the zinc before and after the purification process in ppb (10$^{-9}$ g/g) with uncertainties of 25\%.}
\begin{tabular}{lrr}
Element	& Initial zinc & Purified zinc\\
        \hline
Na	&1000	&30\\
Al	&4000	&35\\
Si	&7000	&40\\
K	&3000	&35\\
Ca	&2000	&70\\
V	&$<$ 200	&$<$ 9\\
Cr	&$<$ 200	&$<$ 90\\
Mn	&$<$ 200	&$<$ 90\\
Fe	&30000	&56\\
Co	&$<$ 200	&$<$ 10\\
Ni	&$<$ 300	&$<$ 90\\
Cu	&10000	&110\\
Mo	&$<$ 2000	&4\\
Cd	&20000	&4300\\
Sn	&15000	&$<$ 300\\
Tl	&$<$ 2000	&180\\
Pb	&$<$ 3000	&290\\
Bi	&$<$ 1000	&$<$ 2\\
Th	&$<$ 2000	&$<$ 0.2\\
U	&$<$ 2000	&$<$ 0.2\\     
    \end{tabular}
    \label{tab:1}
\end{table}
As can be seen from the data, the developed purification method is very effective for the entire range of impurity elements. For most elements the concentration was reduced by one to two orders of magnitude. For example, the iron concentration was reduced by about 500 times, whereas potassium was reduced by a factor of 85. Not listed elements have concentrations which are limited by the instrumental detection level.
Cadmium is the contaminant limiting the purity of material – its concentration was reduced only about 5 times with a final concentration of 4 ppm. This can be explained by two reasons. First, zinc and cadmium are similar from the chemical point of view. Second, at the distillation temperature of zinc, about (460 - 470) $^{o}$C, cadmium and zinc have very similar vapor pressures. As a consequence, cadmium transits almost completely into the condensate of zinc. The mass fraction of the final purified zinc was determined as the difference between 100\% and the total content of major impurities (Cd, Fe, Si, Al, Na, K and Ca). The obtained value is higher than 99.999\%, which meets the requirement on the chemical purity of the materials used for the production of high-quality crystals.

\section{Low background measurement with HPGe detector}
\label{Sec:hpge}
\begin{figure}[ht]
    \centering
    \includegraphics[width=0.5\textwidth]{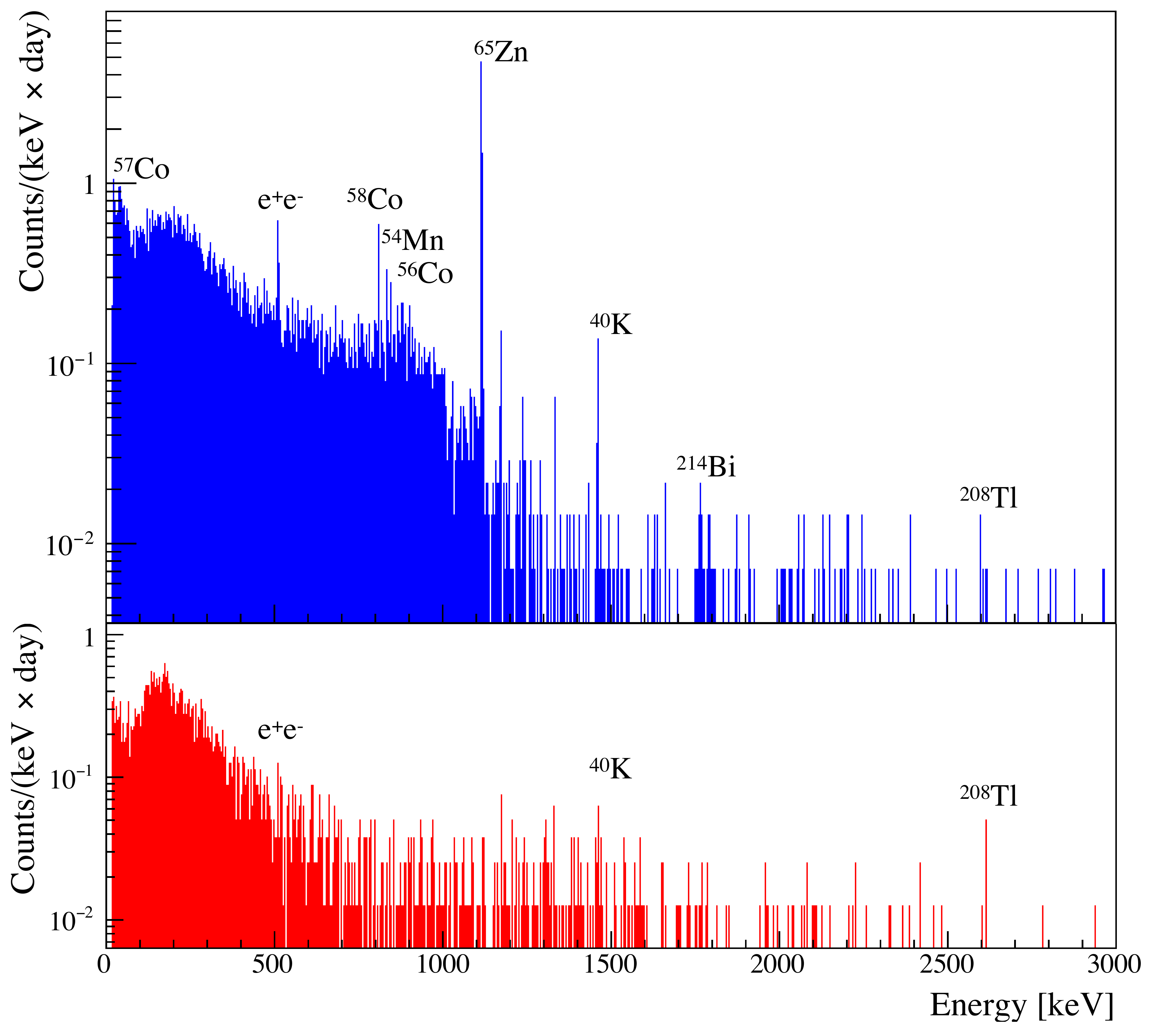}
    \caption{Energy spectrum acquired with the GeMPI-4 ULB-HPGe spectrometer with 10.08 kg of natural zinc over 827.66 hours (top), and without the Zn sample over 474.35 hours (bottom).}
    \label{fig:tot}
\end{figure}

In order to measure the internal radioactive contaminations of the purified granular zinc, we perform a $\gamma$-spectrometry measurement with an ultra-low background high purity germanium detector (ULB-HPGe). We carried out this activity in the STELLA (SubTerranean Low-Level Assay) facility at Gran Sasso National Laboratories (Italy), characterized by an average shielding of 3600 m.~w.~e., details can be found in Refs. \cite{ARPESELLA1996991,NEDER2000191,HEUSSER2006495,2008highly}.
The ULB-HPGe detector is a coaxial p-type germanium detector with an active volume of about 400 cm$^3$ and an optimized design for high counting efficiency in a wide energy range. The energy resolution of the spectrometer is 2.0 keV at the 1332 keV line of $^{60}$Co. To reduce external background, the detector is shielded with a 20 cm layer of low-radioactivity lead, OFHC copper ($\sim$5 cm) and a 5 cm layer of Polyethylene. To prevent radon contamination, the set-up is continuously flushed with high-purity boil-off nitrogen. More details on experimental set-up and detector performance can be found in Ref.~\cite{Laubenstein:2017yjj}. We placed a sample of the purified granular zinc with a mass of 10080 g in a polypropylene container of Marinelli geometry (GA-MA Associates, type 441G) above the end-cap of the ULB-HPGe detector borated. We measured the Zn sample for 827.66 hours, whilst background data was accumulated over 474.35 hours. Fig.~\ref{fig:tot} shows the energy spectra of the two measurements. 
\begin{table}[ht]
    \centering
    \caption{Activity of internal radioactive contaminations in the purified granular zinc measured with a ULB-HPGe detector. Activity values are present in units of mBq/kg, and limits are at 90 \% C.L.}
    \begin{tabular}{lrr}
         Chain	& Nuclide & Activity\\
                &         & [mBq/kg]\\
                \hline
$^{232}$Th	&$^{228}$Ra	&$<$ 0.095\\
            &$^{228}$Th	&$<$ 0.036\\
$^{238}$U	&$^{234}$Th	&$<$ 6.2\\
            &$^{234m}$Pa&$<$ 4.7\\
            &$^{226}$Ra	&$<$ 0.066\\
$^{235}$U	&$^{235}$U	&$<$ 0.091\\
            &           &       \\
            &$^{40}$K	&$<$0.38\\
            &$^{60}$Co	&$<$ 0.036\\
            &$^{137}$Cs	&$<$ 0.033\\
            &{$^{54}$Mn}	&{0.11$\pm$0.02}\\
            &{$^{56}$Co}	&{0.08$\pm$0.02}\\
            &{$^{57}$Co}	&{0.20$\pm$0.09}\\
            &{$^{58}$Co}	&{0.22$\pm$0.04}\\
            &{$^{65}$Zn}	&{5.2$\pm$0.6}\\
    \end{tabular}
    \label{tab:tab2}
\end{table}
The efficiencies for the full-energy absorption peaks used for the quantitative analysis are obtained by Monte-Carlo simulation (code MaGe), based on the GEANT4 software package \cite{5876017}. We obtain values of the limits using the procedure presented in Ref.~\cite{HEISEL2009741}. We report in Tab.~\ref{tab:tab2} the list of internal radioactive nuclides found in the sample.
We found no evidence of any daughter nuclides from the natural decay chains of $^{235}$U, $^{238}$U and $^{232}$Th, so that we set upper limits on the level of less than few mBq/kg. We also report the limits on the activity of other commonly observed nuclides, in particular for $^{40}$K, from natural radioactivity, $^{60}$Co, produced by cosmogenic activation, and artificial $^{137}$Cs.
However, we measure significant activity of some specific isotopes such as $^{54}$Mn, $^{56}$Co, $^{57}$Co, $^{58}$Co and $^{65}$Zn at mBq/kg level. Those isotopes are produced by cosmogenic activation via neutron spallation on naturally occurring zinc isotopes. Among the listed isotopes, $^{54}$Mn is the most long-lived nuclide that decays via electron capture with a Q-value (Q$_{EC}$) of 1377.1 keV and a half-life of 312.12 d. However, the relatively small released energy in this decay makes it not dangerous for exploring of 2$\beta$ decay of the most common isotopes. The same holds for $^{57}$Co (T$_{1/2}$~=~271.79~d, Q$_{EC}$~=~836~keV), $^{58}$Co (T$_{1/2}$~=~70.85~d, Q$_{EC}$~=~2307.4~keV) and $^{65}$Zn (T$_{1/2}$~=~244.26~d, Q$_{EC}$~=~1351.9~keV). $^{56}$Co decays through electron capture with a Q-value of 4566.0 keV, but the relatively short half-life (77.27 d) makes it disappear quickly. These results are confirmed by the CUPID-0 experiment since its ZnSe crystals were made from the zinc measured in this work and the selenium 95\% enriched in $^{82}$Se \cite{Beeman:2015xjv}. Indeed the CUPID-0 background model \cite{Azzolini:2019nmi} attributes to ZnSe crystal 
contaminations much smaller values than the limits reported in this analysis, suggesting that probably during the crystals growth the contaminant segregation further help the purification process. The excellent radiopurity level reached by implementing this technique allowed CUPID-0 to get the lowest background index among cryogenic calorimeters of $3.5\times 10^{-3}$ counts/(keV$\cdot$kg$\cdot$yr)\cite{Azzolini:2019tta}, and several scientific results \cite{Azzolini:2018dyb,Azzolini:2018oph,Azzolini:2019tta,Azzolini:2019yib,Azzolini:2019swx}.

\section{Double beta processes in Zn-64}
\label{Sec:dbd}
Natural zinc contains two potentially 2$\beta$-decaying isotopes, $^{64}$Zn and $^{70}$Zn, whose features are reported in Tab.~\ref{tab:mode}. Since $^{70}$Zn decay would not involve excited states of $^{70}$Ge \cite{PhysRevC.74.024315}, only two electrons sharing the Q$_{\beta\beta}$ energy would be emitted, preventing the decay detection with a HPGe detector. On the contrary, characteristic $\gamma$-rays can be emitted in $\varepsilon\beta^+$ and 2$\varepsilon$ decays of $^{64}$Zn, thus providing distinctive signatures suitable for HPGe spectroscopy.
The CUPID-0 experiment has recently set a new limit $0\nu\varepsilon\beta^+$ of $^{64}$Zn \cite{Azzolini:2020skx}, taking advantage by the ability, offered by the calorimetric approach, to measure the whole $Q_{\varepsilon\beta^+}$ energy. We report in this section a complementary analysis which investigates the decay modes not covered by CUPID-0.
\begin{table}[th]
\centering
\caption{Potentially double beta decaying isotopes of zinc and their features.}
\begin{tabular}{cccc}
Isotope	& Isotopic & Decay mode & Q-value\\
        & abundance [\%]	      &            & [keV]\\
\hline
$^{64}$Zn	& 47.55(0.18)	&$\varepsilon\beta^{+}$/2$\varepsilon$ &	1094.9(0.8)\\
$^{70}$Zn	& 0.68(0.02)	&2$\beta^{-}$	&997.1(2.1)\\
\end{tabular}
\label{tab:mode}
\end{table}
\begin{figure}[t]
    \centering
    \includegraphics[width=0.5\textwidth]{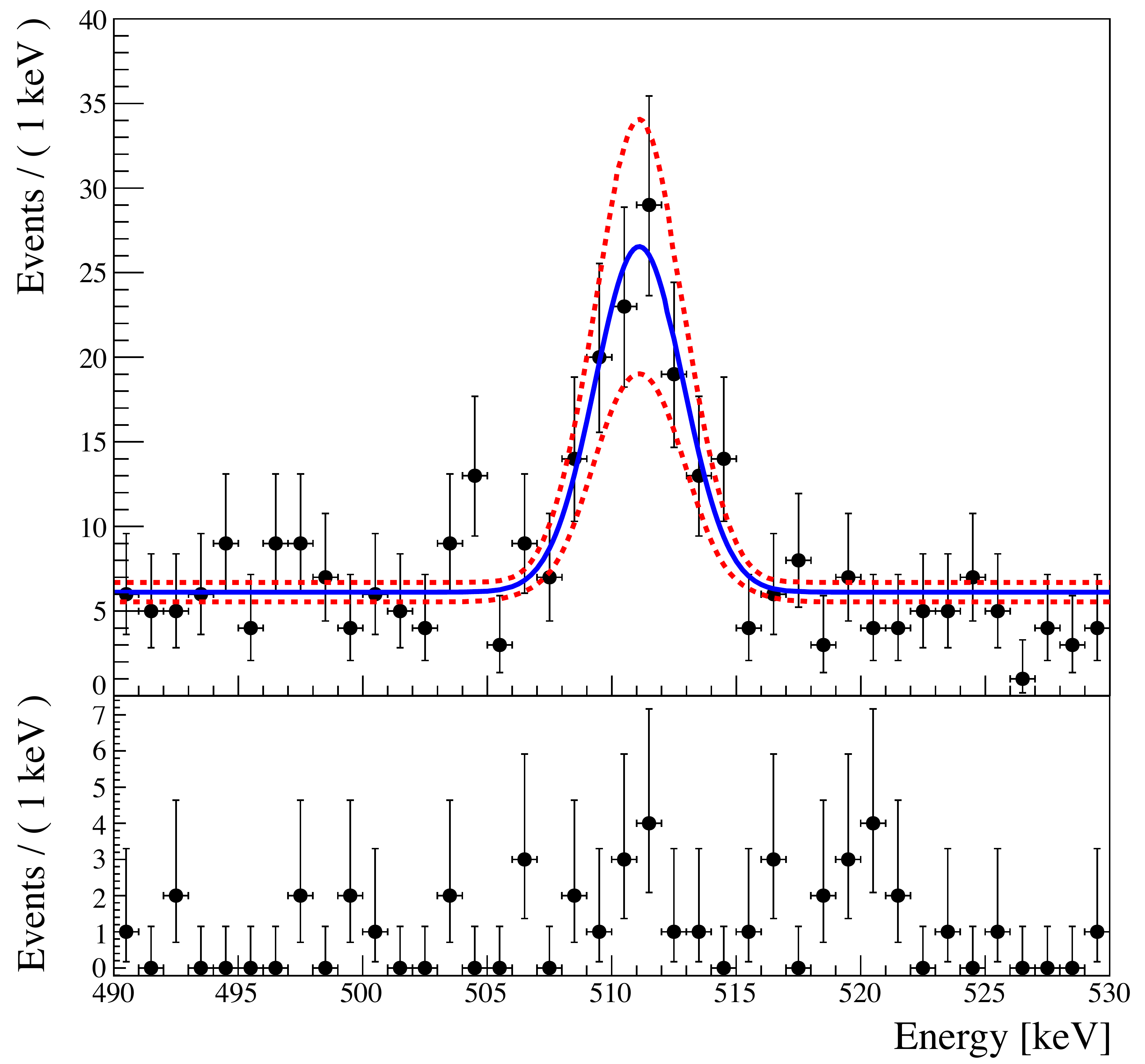}
    \caption{Energy spectra focused on 511 keV peak. Data acquired by the GeMPI-4 ULB-HPGe spectrometer with the zinc sample over 827.66 hours (top), and without sample over 474.35 hours (bottom). The blue line marks the best fit result, while the red dashed lines report a 1~$\sigma$ statistical fluctuation of the number of counts of all the contributions.}
    \label{fig:ECb}
\end{figure}
\subsection{$\varepsilon\beta^{+}$ decay mode}
$^{64}$Zn can decay via electron capture with positron emission:
\begin{equation}
^{64}\text{Zn} + e^- \longrightarrow ^{64}\text{Ni} +  E_{\text{de-exitation}} + e^+~(+2~\nu_e)
\end{equation}
where $E_{\text{de-exitation}}$ is carried by the X-rays or Auger electrons emitted after the capture. In both 0$\nu$ and $2\nu$ cases, the emitted positron generates two 511 keV $\gamma$s, leading to an extra counting rate in the annihilation peak on the total energy spectrum. Since it is impossible to distinguish the two decay modes, we can set a limit for the combination of these two processes. Exploiting a Monte-Carlo simulation, we evaluate the detection efficiency in the experimental setup of a 511 keV $\gamma$-ray, obtaining $\varepsilon = 1.3\%$. Despite the low efficiency, the high $^{64}$Zn isotopic abundance equal to (47.55 $\pm$ 0.18)\% guarantees a large number of emitting isotopes in the sample (N$_{^{64}\text{Zn}}$ = 4.41 $\times$ 10$^{25}$ nuclei), achieving a total exposure of $N_{\text{exp}} = (4.17\pm0.02)\times10^{24}$ emitters$\times$yr. The analysis strategy consists in a Binned Extended Likelihood fit that takes into account all the background sources contributing to the 511 keV peak, selecting as region on interest the energy range [490,530] keV. The model function $\mathcal{F}_{\varepsilon\beta^+}$ is composed of a Gaussian peak over a flat background,
\begin{equation}
\mathcal{F}_{\varepsilon\beta^+} = N_{511}\cdot\mathcal{G}(\mu,\sigma) + N_{\text{b}}\\
\end{equation}
where
\begin{eqnarray}
N_{511}&=& N^{\text{sig}}_{511} + N^{^{65}\text{Zn}}_{511} + N^{^{58}\text{Co}}_{511} + N^{^{56}\text{Co}}_{511}\\
N^{\text{sig}}_{511}&=&\Gamma_{\varepsilon\beta^+}\cdot \varepsilon\cdot N_{\text{exp}}
\end{eqnarray}
The total number of counts of the peak ($N_{511}$) is given by the sum of the expected background events, previously evaluated with a MC simulations assuming the activities in Tab.~\ref{tab:tab2}, and the possible signal ones. The larger contribution into the annihilation peak comes from the decay of $^{65}$Zn with $N^{^{65}\text{Zn}}_{511}=(60 \pm 6)$ counts, while $^{58}$Co and $^{56}$Co produce $N^{^{58}\text{Co}}_{511}=(22 \pm 4)$ and $N^{^{56}\text{Co}}_{511}=(13 \pm 4)$ counts, respectively. 
For all these contributions, we assume a Gaussian prior with the mean and width set to the best estimated values and uncertainties, respectively. The events induced by other isotopes listed in Tab.~\Ref{tab:tab2} are effectively included in the flat background ($N_{\text {b}}$).
The best fit result is $\Gamma_{\varepsilon\beta^{+}} = \bestgeb \times 10^{-22}$ yr$^{-1}$ compatible with zero, thus no significant excess of events was found. We perform a Bayesian analysis integrating the likelihood with a uniform prior and marginalising over the nuisance parameters to set a limit on the half-life of $(0\nu+2\nu)\varepsilon \beta^+$. This is done exploiting the RooStat Bayesian Calculator tools. The resulting upper limit on the decay rate at 90\% C.I. is $\Gamma_{\varepsilon\beta^{+}} < \geb \times 10^{-22}~\text{yr}^{-1}$, corresponding to a lower-limit on the half-life,
\begin{equation}
T_{1/2}^{\varepsilon\beta^{+}} > \teb \times 10^{21}~\text{yr~} \quad \text{at 90\% C.I.}
\end{equation}
This result is almost a factor three better than the previous limit on the half-life of $2\nu\varepsilon\beta^+$, equal to $T_{1/2} > 9.4 \times 10^{20}$ at 90\% C.L. \cite{Belli_2011}. For the $0\nu\varepsilon\beta^+$ a more stringent limit was set by CUPID-0 equal to $1.2\times10^{22}$ yr at 90\% C.I. \cite{Azzolini:2020skx}.
\subsection{2$\varepsilon$ decay mode}
The $^{64}$Zn can decay via double electron capture into the ground state of $^{64}$Ni with the same Q-value as of the $\varepsilon\beta^{+}$ mode, (1094.9 $\pm$ 0.8) keV:
\begin{equation}
^{64}\text{Zn} + 2~e^- \longrightarrow~^{64}\text{Ni} + \gamma_\text{de-exitation}~(+2~\nu_e)
\end{equation}
The $2\nu2\varepsilon$ capture is not considered here because in this case only low energy X-rays are emitted (E = 0.9 - 8.3 keV), which are effectively absorbed by the Zn sample itself. On the other hand, when a $0\nu2\varepsilon$ occurs, we can assume X-rays are accompanied by a bremsstrahlung $\gamma$-quantum with $E_{\gamma} = Q - E_{b1} - E_{b2}$ \cite{Winter:1955zz,Danevich:2020ywz,BARABASH2020121697}, where $E_{b1}$ and $E_{b2}$ are the binding energy of two captured electrons on the corresponding atomic shells of $^{64}$Ni. For Ni atoms, the binding energies on the K and L1 shells are equal to $E_K = 8.3 $ keV and $E_{L1} = 1.01$ keV, respectively. 
\begin{figure}[!ht]
    \centering
    \includegraphics[width=0.5\textwidth]{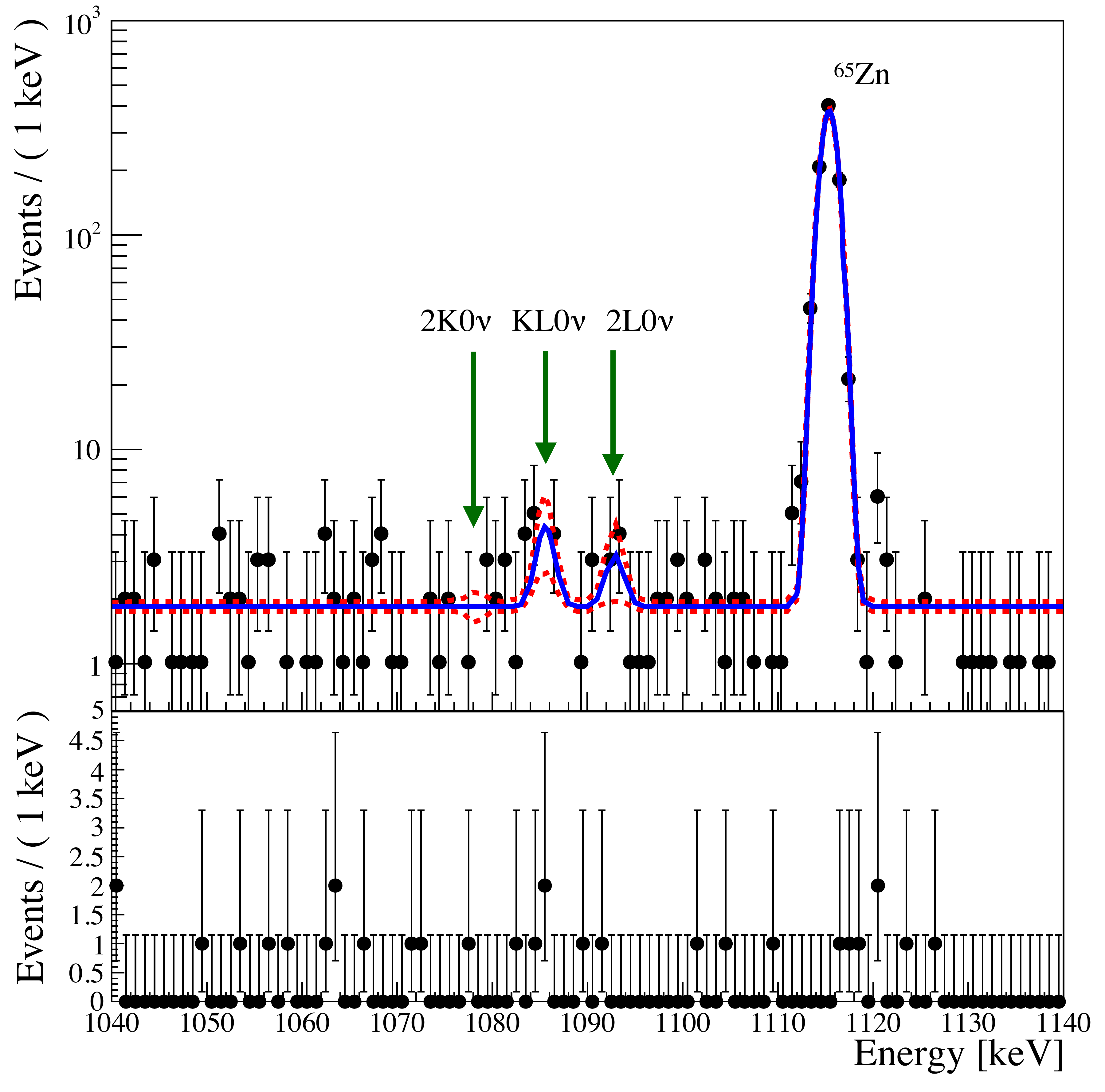}
    \caption{Energy spectrum focused on $0\nu2\varepsilon$ ROI. The peak at 1115 keV is given by the $^{65}$Zn decay. The blue line marks the best fit result, while the red dashed lines report a 1$\sigma$ statistical fluctuation of the number of counts of all the contributions.}
    \label{fig:ECEC}
\end{figure}
Thus, the expected energy of the $\gamma$s for the 2$\varepsilon$0$\nu$ capture $^{64}$Zn to ground state of $^{64}$Ni is equal to: (i) $E_{1} = 1079.0$ keV for 2K0$\nu$; (ii) $E_{2} = 1086.4$ keV for KL0$\nu$; and (iii) $E_{3} = 1093.7$ keV for 2L0$\nu$ processes. 
As in the previous case, a Monte Carlo simulation provides the detection efficiency for gammas of these energies, corresponding to $\varepsilon = 1.1\%$, while the exposure remains the same with respect to the previous analysis. 
Since the atomic mass difference $^{64}$Zn–$^{64}$Ni is currently known with an uncertainty of 0.8 keV, the positions of the expected $0\nu2\varepsilon$ peaks will have the same uncertainty.
Fig.~\ref{fig:ECEC} shows the energy spectrum acquired with the zinc sample in the range [1040,1140] keV. Also in this case, we perform a Binned Extended Likelihood fit on data. The model function $\mathcal{F}_{0\nu2\varepsilon}$ is composed of a flat background and four Gaussian distributions, three of which represent the expected signal and are centered in $\mu_{i} = E_{i}$ ($i=1,2,3$), while the fourth reproduces the $^{65}$Zn peak at 1115 keV,
\begin{equation}
\begin{split}
    \mathcal{F}_{0\nu2\varepsilon} = \sum_{i=1}^{3} N_{i}\cdot\mathcal{G}(\mu_{i},\sigma) +
    N_{^{65}\text{Zn}}\cdot \mathcal{G}(\mu,\sigma) + N_{\text{b}}
\end{split}
\end{equation}
where
\begin{eqnarray}
&N_{i}&=\Gamma^{i}_{0\nu2\varepsilon}\cdot \varepsilon\cdot N_{\text{exp}}\\
&\Gamma_{0\nu2\varepsilon}&=\sum_{i=1}^{3}\Gamma^{i}_{0\nu2\varepsilon}
\end{eqnarray}
We fix the means of the three signal peaks and successively include their uncertainty in the nuisance parameters with a normal prior. We assume all the peaks have the same energy resolution, which is accurately determined on the $^{65}$Zn peak. Since the relative probability of double electron capture from the K and L shells is not available in literature \cite{Blaum:2020ogl}, we do not apply any constrain on $N_{\text{i}}$, considering the three signatures as independent contributes to the process decay rate $\Gamma_{0\nu2\varepsilon}$. 
%In this case, we do not perform a simultaneous fit, as the background spectrum does not contain useful information (see Fig.~\ref{fig:ECEC} bottom panel). 
We observe an excess of events at the KL0$\nu$ peak, but the statistical significance ($1.6~\sigma$) is not such as to suggest the observation of the decay. Following the same procedure as the previous analysis, we set a limit on the decay rate of $\Gamma_{0\nu2\varepsilon} < \gee \times 10^{-22}~\text{yr}^{-1}$ at 90\% C.I., corresponding to a limit on the half-life of 
\begin{equation}
T_{1/2}^{0\nu2\varepsilon}> \tee \times 10^{21}~\text{yr~} \quad \text{at 90\% C.I.}
\end{equation}
This result surpasses by almost one order of magnitude the previous limit, $T_{1/2}^{0\nu2\varepsilon} > 3.2\times10^{20}$ yr \cite{Belli_2011}.

\section{Conclusions}
\label{Sec:conclusions}
In this work, we established an innovative technique for high-purity zinc production, based on the combination of filtration and distillation to remove volatile and non-volatile contaminants. We produced 15 kg of granular zinc with purity grade more than 99.999\%, which was delivered by land transportation to LNGS, where it was characterized with ICP-MS and HPGe spectroscopy. The achieved chemical purity, with a total concentration of entire contaminants less than 10 ppm, completely meets the requirements for the growth of high-quality large-volume crystals. The presence of naturally occurring radio nuclides is excluded, with measured upper limits of few mBq/kg. The only significant radioactive contamination is on the level of few mBq/kg and is due to cosmogenic isotopes. The excellent radiopurity of the zinc sample allowed us to set the most stringent limits to date on the $^{64}$Zn $2\nu\varepsilon\beta^{+}$ and $0\nu2\varepsilon$ decay mode, resulting in $T_{1/2}^{\varepsilon\beta^{+}} > \teb \times 10^{21}~\text{yr~}$, and $T_{1/2}^{0\nu2\varepsilon}> \tee \times 10^{21}~\text{yr~}$, respectively. Exploiting the advantages of spectroscopy with an external source, we overcome the technical difficulty of the CUPID-0 experiment in studying these decays \cite{Azzolini:2020skx} by carrying out a complementary measurement.

\section{Acknowledgments}
This work was partially supported by the European Research Council (FP7/2007-2013) under contract LUCIFER no. 247115.
\bibliography{main}
\bibliographystyle{spphys} 
\end{document}